\documentstyle[twoside]{article}
\newcount\Mac  \Mac=0  
\newcommand{\ifMac}[2]{\ifnum\Mac=1 #1 \else #2 \fi}
\def\putps(#1,#2)(#3,#4)#5#6{\ifnum\Mac=1 \put(#1,#2){\special{picture #5}}
\else  \put(#3,#4){\includegraphics{#6}} \fi}
\def\Red  {}
\def\Black{}
\def\Blue {}
\newcommand{\riga}[1]{\noalign{\hbox{\parbox{\textwidth}{#1}}}\nonumber}
\def\One{\hbox{1\kern-.24em I}}
\newcommand{\lascia}[1]{}
\newcommand{\GeV}{\,{\rm GeV}}
\newcommand{\eV}{\,{\rm eV}}

\newcommand{\NP}{Nucl. Phys.}
\newcommand{\Jhep}{{\rm J.hep}}
\newcommand{\PRL}{Phys. Rev. Lett.}
\newcommand{\PL}{Phys. Lett.}
\newcommand{\PR}{Phys. Rev.}

\newcommand{\eq}[1]{~(\ref{eq:#1})}

\def\Ord{{\cal O}}  
 
\def\circa#1{\,\raise.3ex\hbox{$#1$\kern-.75em\lower1ex\hbox{$\sim$}}\,}
\makeatletter
\def\art{\@ifnextchar[{\eart}{\oart}}
\def\eart[#1]#2#3#4#5#6{{\rm #2}, {\em #3 \rm #4} {\rm (#6) #5 ({\em #1})}}
\def\hepart[#1]#2{{\rm #2, \em#1}}
\newcommand{\oart}[5]{{\rm #1}, {\em #2 \rm #3} {\rm (#5) #4}}

\newcounter{alphaequation}[equation]
\def\thealphaequation{\theequation\hbox to
0.6em{\hfil\alph{alphaequation}\hfil}}
\def\eqnsystem#1{
\def\@eqnnum{{\rm (\thealphaequation)}}
\def\@@eqncr{\let\@tempa\relax \ifcase\@eqcnt \def\@tempa{& & &} \or
  \def\@tempa{& &}\or \def\@tempa{&}\fi\@tempa
  \if@eqnsw\@eqnnum\refstepcounter{alphaequation}\fi
\global\@eqnswtrue\global\@eqcnt=0\cr}
\refstepcounter{equation} \let\@currentlabel\theequation \def\@tempb{#1}
\ifx\@tempb\empty\else\label{#1}\fi
\refstepcounter{alphaequation}
\let\@currentlabel\thealphaequation
\global\@eqnswtrue\global\@eqcnt=0 \tabskip\@centering\let\\=\@eqncr
$$\halign to \displaywidth\bgroup \@eqnsel\hskip\@centering
$\displaystyle\tabskip\z@{##}$&\global\@eqcnt\@ne
\hskip2\arraycolsep\hfil${##}$\hfil& \global\@eqcnt\tw@\hskip2\arraycolsep
$\displaystyle\tabskip\z@{##}$\hfil
\tabskip\@centering&\llap{##}\tabskip\z@\cr}
\def\endeqnsystem{\@@eqncr\egroup$$\global\@ignoretrue} \makeatother

\begin{document}

\twocolumn[
\centerline{\bf OUTP--99--30P \hfill    IFUP--TH/32--99}
\centerline{\bf hep-ph/9906470 \hfill SNS--PH/99--10} \vspace{1cm}
\centerline{\LARGE\bf\Red Vacuum oscillations of quasi degenerate solar neutrinos\footnotemark[1]}

\bigskip\bigskip\Black
\centerline{\large\bf Riccardo Barbieri} \vspace{0.3cm}
\centerline{\em Scuola Normale Superiore and INFN, sezione di Pisa, I-56126 Pisa, Italia}\vspace{0.3cm}
\centerline{\large\bf Graham G.\ Ross} \vspace{0.3cm}
\centerline{\em Department of Physics, University of Oxford, 1 Keble Road, Oxford  OX1 3NP, UK}\vspace{0.3cm}
\centerline{\large  and }\vspace{0.3cm}
\centerline{\large\bf Alessandro Strumia}\vspace{0.3cm}
\centerline{\em Dipartimento di fisica, Universit\`a di Pisa and INFN,
sezione di Pisa, I-56126 Pisa, Italia}\vspace{0.3cm}

\bigskip\bigskip\Blue

\centerline{\large\bf Abstract}
\begin{quote}\large\indent
The atmospheric neutrino oscillations and the vacuum oscillation solution
of the solar neutrino problem can be consistently described by a doubly or triply 
degenerate neutrino spectrum as long as the high level of degeneracy
required is not spoiled by radiative corrections. We show that this is the case for
neutrino mass matrices generated by symmetries. This imposes a strong constraint
on the mixing angles and requires the mixing should be close to bi-maximal. We briefly 
discuss the relevance of our results for the measurability of the neutrino
spectrum.
\end{quote}\Black
\vspace{1cm}] 

\section{Introduction}\footnotetext[1]{Research supported in part by the EEC under TMR contract ERBFMRX-CT96-0090.}
Observations of atmospheric and solar neutrinos provide very significant
indications that neutrinos oscillate between different mass eigenstates $%
m_{i}$~\cite{sunexp,atmexp}. As a result, progress has been made in
experimentally determining these masses as well as the related mixing
parameters. In what follows we assume just 3 light Majorana neutrinos and we
use a standard notation for the leptonic mixing matrix 
\begin{equation}
V=R_{23}(\theta _{23})\mathop{\rm diag}(1,e^{i\phi },1)R_{13}(\theta
_{13})R_{12}(\theta _{12})  \label{eq:Vunitary}
\end{equation}
where $R_{ij}(\theta _{ij})$ represents a rotation by $\theta _{ij}$ in the $%
ij$ plane. Within this framework and with this notation, the present
situation can be summarised as follows:
\begin{itemize}
\item It is very likely, although still awaiting confirmation, that 
\begin{eqnarray}\nonumber
&&\Delta m_{23}^2\equiv m_{{\rm atm}}^2=10^{-(3\div 2)}\,{\rm eV}^2, \\
&&\sin^22\theta _{23}\circa{>}0.8  \label{eq:atm}
\end{eqnarray}
\item 
It is not unlikely that 
\begin{eqnarray}\nonumber
&&\Delta m_{12}^2\equiv m_{{\rm sun}}^2\circa{<}10^{-4}\,{\rm eV}^2,\\
&&\sin^2\theta _{13}\circa{<}0.1  \label{eq:sun1}
\end{eqnarray}
with~\cite{sunfit}
\begin{eqnsystem}{sys:sun}\nonumber
&&3\cdot10^{-6}\eV^2\circa{<}m_{\rm sun}^2\circa{<}10^{-5}\eV^2,\\ 
&&2\cdot10^{-3}<\sin^2 2\theta_{12}<2\cdot 10^{-2}\\
\riga{\hspace{1cm}or}\\ \nonumber
&&10^{-5}\eV^2\circa{<}m_{\rm sun}^2\circa{<}10^{-4}\eV^2,\\ 
&&0.6\circa{<}\sin^2 2\theta_{12}<0.95\\
\riga{\hspace{1cm}or}\\ \nonumber
&&5\cdot 10^{-11}\eV^2\circa{<}m_{\rm sun}^2\circa{<}10^{-9}\eV^2,\\ 
&&\sin^2 2\theta_{12}\circa{>}0.6
\end{eqnsystem}
corresponding respectively to the small angle MSW (SAMSW), large angle MSW
(LAMSW) or vacuum oscillation (VO) solutions of the solar neutrino problem.

\end{itemize}
It is clearly of great importance to confirm or disprove this picture and further
constrain the allowed range of the parameters.

Even accepting~(\ref{eq:atm},\ref{eq:sun1},\ref{sys:sun}), which we do
hereafter, the neutrino spectrum is not determined. As is well known, three
different possibilities exist:

\begin{enumerate}
\item  ``degenerate'': 
$$
m_{1}\approx m_{2}\approx m_{3}\,\raise.3ex\hbox{$>$\kern-.75em\lower1ex%
\hbox{$\sim$}}\,m_{{\rm atm}} 
$$

\item  ``pseudo-Dirac'':
$$
m_{1}\approx m_{2}\approx m_{{\rm atm}}\gg m_{3} 
$$

\item  ``hierarchical'':
$$
m_{3}\approx m_{{\rm atm}}\gg m_{2}\approx m_{{\rm sun}}\,\raise.3ex%
\hbox{$>$\kern-.75em\lower1ex\hbox{$\sim$}}\,m_{1} 
$$
\end{enumerate}
It is extremely important to know, with a minimum of theoretical bias, which
spectrum is realized in nature. However, at the moment we only know that
the heaviest neutrino weights less than a few eVs from direct $\beta $-decay
searches or from astrophysical and cosmological data.
Three different areas of experimental developments can have an impact on
this issue\footnote{%
We remind that in this paper we are considering the case there
are just three light neutrinos. Of course our analysis would require
revision if this proves not to be the case.}:

\begin{enumerate}
\item  The determination of the actual solution 
of the solar neutrino problem.

\item  The neutrinoless double-beta decay searches.

\item  The cosmological signals of a neutrino rest mass.
\end{enumerate}
Prior to the discussion of the experimental potential in this area there is,
however, one relevant theoretical problem. The VO of solar neutrinos can be
consistently described by the `degenerate' or the `pseudo-Dirac' spectra
only as long as the high level of degeneracy required is not spoiled by
radiative corrections~\cite{EL,madrid}. We investigate this question in this
paper, concentrating on the form of neutrino mass matrices motivated by
symmetries, as previously suggested~\cite{lungo,PD,deg}.
These mixing matrices are all characterized by having $\theta_{12}=\pi/4$ and $\theta_{13}=0$.
We will show that
these mass matrices have a sufficient degree of stability against radiative
corrections to make the VO solution consistent with both doubly and triply
degenerate neutrino spectra.
In conjunction with the experimental requirement that $\theta_{23}$ is
approximately $\pi/4$, this
leads to the so called bimaximal mixing matrix~\cite{bimax}.
Similar conclusions have also been reached in~\cite{CEIN,Wu}.
Finally we will consider the necessary and sometimes
sufficient conditions needed to determine the full neutrino spectrum.

\section{Radiative corrections to the neutrino mass textures}
We concentrate on neutrino mass matrices, in the charged
lepton flavour basis, of the form 
\begin{equation}
M_{\nu }^{0}=m~R_{23}(\theta _{23})\pmatrix{0&1&0\cr 1&0&0\cr 0&0&z}%
R_{23}^{T}(\theta _{23})  \label{eq:11z}
\end{equation}
where: i) for the `degenerate' case, $m$ is the common neutrino mass, $m\,%
\raise.3ex\hbox{$>$\kern-.75em\lower1ex\hbox{$\sim$}}\,m_{{\rm atm}}$, and $%
z=e^{i\phi }(1+\delta )$ with $2\delta +\delta^2=(m_{{\rm atm}}/m)^2$;
ii) for the `pseudo-Dirac' case, $m=m_{{\rm atm}}$ and $|z|$ is negligibly
small.

In both cases the small splitting necessary to describe the VO of solar
neutrinos is neglected. It could come from an explicit extra term in\eq{11z}
or it could even be generated by the same radiative correction
effects that we are going to discuss. $R_{23}(\theta _{23})$ is the large
angle rotation in the 23 sector that accounts for atmospheric neutrino
oscillations. As noted above the neutrino mass matrix~(\ref{eq:11z}) is in
the flavour basis, i.e.\ it is associated with diagonal charged lepton mass
matrices. For the `degenerate' case, texture~(\ref{eq:11z}) was motivated
in~\cite{deg} on the basis of a spontaneously broken SO(3) flavour
symmetry. For the `pseudo-Dirac' case, $z=0$, texture~(\ref{eq:11z}) was
introduced in~\cite{lungo,PD} as a consequence of an unbroken abelian
symmetry, $L_{e}-L_{\mu }-L_{\tau }$. The very fact that this last symmetry
is compatible with the Yukawa couplings (the charged fermion masses) of the
Standard Model or of the Minimal Supersymmetric Standard Model makes it
clear that radiative corrections will not destabilize~(\ref{eq:11z}) in
either case. The issue is more tricky in the fully `degenerate' case, with $%
|z|\approx 1$, since~(\ref{eq:11z}) is obtained, together with a diagonal
charge lepton matrix, only after spontaneous symmetry breaking of the SO(3)
symmetry with scalar vacuum expectation values in appropriate directions~\cite{deg}.
Note that~(\ref{eq:11z}), which must be viewed as an initial
condition valid at some scale $\Lambda $, can be rewritten as 
\begin{equation}
M_{\nu }^{0}=m~V^{\ast }\cdot \mathop{\rm diag}(-1,1,z) \cdot V^{\dagger }
\end{equation}
with 
\[
V=R_{23}(\theta _{23})\cdot R_{12}(\pi /4).
\]
In full generality, up to universal corrections and negligibly small effects
of the muon and electron Yukawa couplings, the renormalized neutrino mass at
a scale $\mu $ below $\Lambda $ is given in logarithmic approximation by~\cite{EL,madrid} 
\begin{eqnarray}
M_{\nu } &=&I_{\tau }\cdot M_{\nu }^{0}\cdot I_{\tau }  \label{eq:MnuRGE} \\
I_{\tau } &=&\mathop{\rm diag}(1,1,1+\epsilon )  \nonumber \\
\riga{where}\\
\epsilon  &=&\frac{g_{\tau }^2}{(4\pi )^2}\ln \frac{\Lambda }{\mu }\{%
\frac{1}{2},\frac{-1}{\cos^2\beta }\}  \nonumber
\end{eqnarray}
and $g_{\tau }=m_{\tau }/v\approx 0.01$ is the $\tau $ Yukawa coupling in the SM.
The two factors in parenthesis stand for the SM or for the MSSM
contributions respectively, with $\tan \beta =v_{2}/v_{1}$ being the usual
parameter related to the ratio of the Higgs vevs. For the purposes of this
discussion it is sufficient to take $\mu =M_{Z},$ ignoring the small
corrections due to the different possible thresholds at the electroweak
scale. Since $(g_{\tau }/4\pi )^2\approx 0.6~10^{-6}$, $\epsilon $ is
significantly larger, for any value of $\Lambda $ and $\tan \beta $, than
the relative splitting needed to account for the VO solution of solar
neutrinos, 
\begin{equation}
\frac{\Delta m}{m}\leq  \frac{m^2_{{\rm sun}}}{m^2_{{\rm atm}}}\circa{<}10^{-7}.
\end{equation}
Thus if $\Delta m/m={\cal O}(\epsilon )$ there will be a conflict with the
required level of degeneracy of the renormalized neutrino masses in~(\ref{eq:MnuRGE})~\cite{EL,madrid}.

We will demonstrate that for the neutrino mass matrices of the form given in
(\ref{eq:11z}), due to the underlying symmetries, in fact
${\Delta m}/{m}={\cal O}(\epsilon^2)$
and hence the vacuum oscillation solution is quite stable
against radiative corrections. It is most convenient to discuss the
symmetries in the basis rotated by $\theta _{23}$ in which the neutrino mass
matrix has the form 
\begin{equation}
\tilde{M}_{\nu }^{0}=R_{23}^{T}(\theta _{23})M_{\nu }^{0}R_{23}(\theta
_{23})=m\pmatrix{0&1&0\cr 1&0&0\cr 0&0&z}  \label{new1}
\end{equation}
This is invariant under a ${\rm U}(1)$ rotation under which the states $1,2$ and 
$3$ have charges $+1,$ $-1$ and $0$ respectively. Now consider the effect of
the radiative corrections. If they preserve the ${\rm U}(1)$ they will leave the
zero structure of the mass matrix intact and in turn this leaves one
degenerate pair of neutrinos. It is useful to rewrite~(\ref{eq:MnuRGE}) in
terms of the matrix $\tilde{I}_{\tau }$ defined by 
\[
\tilde{I}_{\tau }=R_{23}^{T}I_{\tau }R_{23}=\One +\epsilon X
\]
where
\begin{equation}
X=\left( 
\begin{array}{ccc}
0 & 0 & 0 \\ 
0 & s^2_{23} & c_{23}s_{23} \\ 
0 & c_{23}s_{23} & c^2_{23}
\end{array}
\right)   \label{eq:new2}
\end{equation}
and $c_{ij}\equiv \cos \theta _{ij}$, $s_{ij}\equiv \sin \theta _{ij}$.
In terms of $\tilde{I}_{\tau }$ the renormalized mass matrix in the same
basis is given by 
\begin{equation}
\tilde{M}_{\nu }=\tilde{I}_{\tau }.\tilde{M}_{\nu }^{0}.%
\tilde{I}_{\tau }  \label{eq:new3}
\end{equation}
Let us consider the order at which the degeneracy of the light neutrinos is
lifted. Since $\tilde{I}_{\tau }$ comes from wave function
renormalization, only its diagonal elements are invariant under the ${\rm U}(1)$
discussed above. Thus ${\rm U}(1)$ breaking effects arise through the elements $X_{23},$ $X_{32}$.
But these matrix elements, being off-diagonal, would
remove the degeneracy of the eigenvalues at $\Ord(\epsilon)$
only if $\epsilon\circa{>}|z-1|$, which is not the case,
as required by the atmospheric neutrino anomaly.
\lascia{
At leading order in $\epsilon $ we have 
\begin{equation}
\tilde{M}_{\nu }=\tilde{M}_{\nu }^{0}+\epsilon (\tilde{M}_{\nu
}^{0}\cdot X+X\cdot \tilde{M}_{\nu }^{0})  \label{new4}
\end{equation}
Only the ${\cal O}(\epsilon )$ corrections to the diagonal $(1,2)$ and $(2,1)$
matrix elements can give rise to ${\cal O}(\epsilon )$ corrections to
the eigenvalues of the mass matrix\footnote{This is true provided the third neutrino is not
exactly degenerate with the other
two, as required by the atmospheric neutrino anomaly.}. From (\ref{new4}) one may see that the symmetry
breaking elements of $X$ do not contribute to these elements at ${\cal O}(\epsilon )
$ and hence the mass matrix structure following from the ${\rm U}(1)$ remains. As
a result the degeneracy of the two light neutrinos is maintained at this
order, the ${\cal O}(\epsilon )$ terms simply changing the magnitude of the
degenerate mass and that of the third neutrino. At ${\cal O}(\epsilon^2)$ the
symmetry breaking elements $X_{23},$ $X_{32}$ do contribute and we expect
the degeneracy to be broken at this order.}

\section{Discussion of the results}
We turn now to the quantitative determination of these effects. To do this
is useful to rewrite (\ref{eq:MnuRGE}) in terms of the matrix 
\begin{equation}
I_{\tau }^{\prime }=V^{\dagger }I_{\tau }V
\end{equation}
as 
\begin{equation}
M_{\nu }=V^{\ast }\cdot I_{\tau }^{\prime T}M_{{\rm diag}}^{0}I_{\tau
}^{\prime }\cdot V^{\dagger }=V^{\ast }\cdot M_{\nu }^{\prime }\cdot
V^{\dagger }  \label{eq:MIR}
\end{equation}
where, by explicit calculation, 
\begin{equation}
\frac{M_{\nu }^{\prime }}{m}=\pmatrix{-1-\epsilon s^2_{23} &0&\epsilon'(-1+z)/4\cr 0 &
1+\epsilon s^2_{23} &\epsilon'(-1-z)/4\cr \epsilon'(-1+z)/4 &
\epsilon'(-1-z)/4 & z(1+2\epsilon c_{23}^2)},  \label{eq:M'IR}
\end{equation}
$\epsilon^{\prime }\equiv \sqrt{2}\epsilon \sin 2\theta _{23}$.
Eqs.~(\ref{eq:MIR}) and~(\ref{eq:M'IR}) are the basic expressions for the renormalized
neutrino mass matrix in the flavour basis.
{
The renormalization of the `pseudo-Dirac' spectrum is immediately obtained by setting $z=0$ in~(\ref
{eq:M'IR}). This gives
\begin{equation}
M_{\nu }(z=0)=m~R_{23}(\theta _{23})\cdot
\end{equation}
\[
\cdot 
\pmatrix{0&1+\epsilon s_{23}^2 &-\epsilon s_{23}c_{23}\cr
1+\epsilon s_{23}^2 &0&0\cr
-\epsilon s_{23}c_{23}&0&0}\cdot R_{23}^{T}(\theta _{23}) 
\]
which keeps the original form, as anticipated by our symmetry arguments,
with (small) renormalizations of the angle $\theta _{23}$ and of the overall
scale $m$.

\medskip

For the degenerate case, $z=e^{i\phi }(1+\delta ),$ it is best to work with the
hermitian matrix 
\begin{equation}
M_{\nu }M_{\nu }^{\dagger }=V^{\ast }M_{\nu }^{\prime }M_{\nu
}^{\prime \dagger }V^{T}
\end{equation}
where, from~(\ref{eq:M'IR}) 
\begin{equation}
M_{\nu }^{\prime }M_{\nu }^{\prime \dagger }=m^{\prime 2}\pmatrix{1&0&i%
\epsilon'e^{-i\phi/2}\sin\frac{\phi}{2}\cr 0&1&-\epsilon' e^{-i\phi
/2}\cos\frac{\phi}{2}\cr \hbox{h.c.} & \hbox{h.c.} &1+2\delta}  \label{eq:MM}
\end{equation}
with $m^{\prime }=m(1+\epsilon s_{23}^2)$ and up to irrelevant
corrections. The eigenvalues of this matrix, i.e.\ the renormalized squared
neutrino masses are 
\begin{eqnarray}
m_{1}^2 &=&m^{\prime 2}  \nonumber \\
m_{2}^2 &=&m^{\prime 2}(1+\delta -(\delta^2+\epsilon^{\prime 2})^{1/2})
\nonumber \\
&\approx &m^{\prime 2}(1-\epsilon^{\prime 2}/2\delta )  \nonumber \\
m_{3}^2 &\approx &m^{\prime 2}(1+2\delta ).  \label{ev}
\end{eqnarray}
As anticipated by our symmetry argument the degeneracy between the light
states is lifted at ${\cal O}(\epsilon^2).$ Note that had we dropped $\delta $
in $m_{2}^2$ the correction would have been of order $\epsilon$.
The inclusion of the atmospheric mass splitting is crucial and explains why our conclusions
about the compatibility of a `degenerate' neutrino spectrum with VO solar oscillations
differ from~\cite{EL,madrid}.

To the extent~(\ref{eq:11z}) represents the exact initial condition for $%
M_{\nu }$, these renormalized eigenvalues would give\footnote{%
The signs of the mass squared splittings, $ m_{{\rm sun}}^2$ and $ m_{{\rm atm}}^2$, are
irrelevant.} 
\begin{equation}
\frac{m_{{\rm sun}}^2}{m^2}=\frac{\epsilon^2}{\delta }\sin
^22\theta _{23},\qquad \frac{m_{{\rm atm}}^2}{m^2}=2\delta 
\end{equation}
i.e. 
\begin{equation}
\Blue m_{{\rm sun}}^2m_{{\rm atm}}^2=2m^{4}\epsilon^2~\sin
^22\theta _{23}  \label{eq:pred}
\end{equation}
More generally there could be a splitting of the original unrenormalized
eigenvalues, which makes~(\ref{eq:pred}) a rough estimate of a lower bound
on $ m_{{\rm sun}}^2 m_{{\rm atm}}^2$, barring strong
accidental cancellations. Numerically, choosing $\Lambda =10^{5\div 16}\,{\rm GeV}$\footnote{For simplicity,
we are neglecting the running of the $\tau$ Yukawa coupling.
In the MSSM with moderate $\tan\beta$ and for $\Lambda\approx 2~10^{16}\GeV$,
our approximation is $\sim2$ times larger than the exact result.}
\begin{equation}
\frac{m_{{\rm sun}}^2m_{{\rm atm}}^2}{(1\div 20)10^{-11}\,{\rm eV}^4}%
\,\raise.3ex\hbox{$>$\kern-.75em\lower1ex\hbox{$\sim$}}\,
\left( \frac{m}{\,{\rm eV}}\right)^{4}\{1,(\frac{2}{\cos^2\beta })^2\}  \label{eq:Npred}
\end{equation}
to be compared with 
\begin{equation}  \label{eq:Nexp}
m_{{\rm sun}}^2m_{{\rm atm}}^2|_{{\rm exp}}=10^{-(11\div 14)}\,{\rm eV}^4.
\end{equation}
Eq.~(\ref{eq:Npred}) is our main result and clearly shows that even a
threefold degenerate spectrum, with the neutrino mass matrix following from
an underlying symmetry, is compatible with the VO solution of the solar
neutrino problem. The radiative correction due to the $\tau $ Yukawa
coupling is actually a candidate for generating the VO $ m_{{\rm sun}}^2$ splitting.
If this is the case, i.e.\ the bound in~(\ref{eq:Npred})
is saturated, a reduction of the uncertainty in the right handed side of~(\ref{eq:Nexp})
would allow a rather precise determination of the average
neutrino mass $m.$ At present the lower bound for $m$ is given by $m_{{\rm %
atm}}=(0.03\div 0.1)\,{\rm eV}$. The upper bound follows from~(\ref{eq:Npred}).
Note that values of cosmological interest, $\sum_{\nu }m_{\nu }\sim {\rm eV}$, cannot be
safely excluded on the basis of~(\ref{eq:Npred}).

\medskip

It is of interest to note that
$\theta _{12}=\pi /4$ and $\theta_{13}=0$ have been both necessary to avoid corrections of order $\epsilon $ to
$m_{{\rm sun}}^2$ that would drastically change our conclusions
(on the contrary the value of $\theta _{23}$ does not crucially affect the magnitude of the radiative corrections).
At first sight it could appear that
the complex 12 rotation
necessary to re-diagonalize the RGE-corrected mass matrix\eq{MM}
$$U_{12}(\phi/2 )=\mathop{\rm diag}(i,1,1)R_{12}(\phi/2 )\mathop{\rm diag}(-i,1,1)$$
induces a too large renormalization of $\theta _{12}$
unless the phase $\phi$ is very small.
This would be a problem for the model in~\cite{deg}, where $\phi\approx\pi/2$.
However this complex rotation does not affect $\theta_{12}=\pi /4$,
as may be shown by means of the identity
\[
U_{12}(\phi /2)R_{12}(\pi /4)=R_{12}(\pi /4)\mathop{\rm diag}(e^{i\phi
/2},e^{-i\phi /2},1).
\]
Consequently the small RGE effects only induce small
RGE corrections, of order $\epsilon/\delta$, to the
initial values $\theta _{12}=\pi /4$ and $\theta _{13}=0$ of
the mixing angles even if $\phi$ is large\footnotemark[1]\footnotetext[1]{Radiative corrections to the mixing angles
in presence of two degenerate neutrinos
have also been studied in~\cite{oka,madrid,CEIN,CKP}.
The fact that, with a `pseudo-Dirac' spectrum and
appropriate correlations between the mixing angles, a cancellation of the $\Ord(\epsilon)$ corrections
takes place has been also observed in~\cite{CEIN}.
Although these delicate cancellations happen in very narrow regions of the mixing angles
$$|\theta_{13}|<m_{{\rm sun}}^2/4\epsilon m^2,\qquad 
|\theta _{12}-\pi/4|<m_{{\rm sun}}^2/4\epsilon m^2$$
we consider this situation of physical interest because
mixing parameters inside these narrow regions are motivated by symmetries~\cite{lungo,PD,deg}.}.

Given that experiments indicate a large $\theta _{23}\approx \pi/4$ and
disfavour a large $\theta_{13}$,
we conclude that the degenerate case generating a VO solution to
the solar neutrino problem, stable against radiative corrections, must lie
very close to the bimaximal mixing solution\footnotemark[2]\footnotetext[2]{We have shown
that the radiative corrections to the solar splitting vanish at leading order in $\lambda_\tau$
if $V=R_{23}(\theta _{23})\cdot R_{13}(0)\cdot R_{12}(\pi /4)$.
The same thing happens for a more general $V=R_{12}(\Delta \theta_{12}) \cdot R_{23}(\theta_{23})R_{12}(\pi/4)$,
since the rotation $R_{12}(\Delta \theta_{12})$ is irrelevant
as long as the $\mu$ and $e$ Yukawa couplings are neglected.
It seems not unconceivable that even a $V$ of this form might result as a consequence
of an approximate symmetry,
although this is not the case in the models in~\cite{PD,deg}.
When rewritten in the standard parametrization\eq{Vunitary} $V$ corresponds
to having $\theta_{13}\neq0$ with
$\sin \theta_{13}=\tan\theta_{23}\cdot \tan(\theta_{12}-\pi/4)$,
a relation equivalent to eq.~(19) in~\cite{CEIN}.
The apparently different correlation presented in eq.~(35) of~\cite{CKP}
is an equivalent parametrization of the same $V$.}.

\medskip

Up to now we have concentrated on the VO solution.
Our analysis immediately applies to the SAM\-SW and LAMSW cases as well.
In such cases, however, the required level of degeneracy
is not incompatible even with a splitting of $\Ord(\epsilon)$.
Hence no significant restriction on the mixing parameters arises.

In conclusion, all the three favorite oscillation solutions of the solar neutrino problem
(SAMSW, LAMSW or VO) are compatible with all the three possible
spectra of neutrinos (`degenerate', `pseudo-Dirac', or `hierarchical').
The eventual identification of VO as the true solution of the solar neutrino problem
would not imply a hierarchical spectrum of neutrinos.
Even with a triply degenerate spectrum, the known radiative corrections are not too large
if the mixing angles have certain values motivated by symmetries.
In the next section we discuss how the true neutrino spectrum could be 
identified by conceivable experiments.


\section{Will the neutrino spectrum ever be measured?}
The conclusions of the previous section make even more acute the problem of the possible experimental determination
of the neutrino spectrum~\cite{BW}.

If $\theta_{13}$ is non-zero, due to matter effects, the sign of the atmospheric mass
splitting might be measurable by the study of $\nu_\mu\to\nu_e$ and $\bar{\nu}_\mu\to\bar{\nu}_e$ oscillations
in a long baseline experiment~\cite{Lipari},
using a $\nu$ beam generated by a neutrino factory.
In turn this would allow to discriminate between a `hierarchical' spectrum 
(where $m_3^2\gg m_{1,2}^2$)
and a `pseudo-Dirac' spectrum 
(where $m_{1,2}^2\gg m_3^2$).}

The $0\nu ,2\beta $ decay searches have set a strong constraint on the
modulus of the relevant element of the neutrino mass matrix 
\begin{equation}
|M_{\nu }|_{ee}=|c_{13}(c_{12}^2m_{1}+s_{12}^2m_{2}e^{2i\varphi
_{2}}+s_{13}^2m_{3}e^{2i\varphi _{3}})|  \label{eq:Mnuee}
\end{equation}
where $\varphi _{i}$ are arbitrary phases At the moment, taking into account
the uncertainty on the nuclear matrix element, it is $|M_{\nu
}|_{ee}<(0.2\div 0.4)\,{\rm eV}$~\cite{betabetaNow}. The sensitivity of $%
0\nu ,2\beta $ experiments is thought to be improvable by about one order of
magnitude~\cite{betabeta}.
A signal for a neutrino mass might also be obtained from studies of
large scale structures in the universe, together with accurate
measurements of anisotropies in the temperature of the Cosmic Background Radiation.
With the standard cosmological model as reference paradigm,
a sensitivity to a total neutrino mass $\sum_\nu m_\nu \circa{>} 0.3\eV$ is thought
to be achievable~\cite{cosm}.
The impact of these searches on the issue under
consideration can be summarized as follows, as explained below:
\begin{enumerate}
\item  Finding a $0\nu ,2\beta $ and/or a cosmological signal, at the level
specified above, would prove the `degenerate' or the `pseudo-Dirac' spectrum.
Different signals can discriminate between `degenerate' or `pseudo-Dirac' 
spectra and/or imply a specific solution of the
solar neutrino problem.

\item  If neither $0\nu ,2\beta $ or a cosmological signal will be found, further
progress will require knowing that $|M_{\nu }|_{ee}<m_{{\rm atm}}$.
In such a case, establishing SAMSW for solar neutrinos will prove the `hierarchical'
spectrum. On the contrary, LAMSW or VO for solar neutrinos would not allow
any straightforward conclusion on the spe\-ctrum itself.
\end{enumerate}
Finding $|M_\nu|_{ee}>0.01\,{\rm eV}$ and/or $\sum_\nu m_\nu>0.3 \,{\rm eV}$
would be against the `hierarchical' spectrum since, in such a case 
\[
|M_\nu|_{ee}\le |s_{13}^2 m_{{\rm atm}}+m_{{\rm sun}}|\,\raise.3ex%
\hbox{$<$\kern-.75em\lower1ex\hbox{$\sim$}}\,0.01\,{\rm eV} 
\]
and 
\[
\sum_\nu m_\nu\approx m_{{\rm atm}}\,\raise.3ex\hbox{$<$\kern-.75em\lower1ex%
\hbox{$\sim$}}\,0.1\,{\rm eV} 
\]
upon use of~(\ref{eq:atm},\ref{eq:sun1},\ref{eq:Mnuee}). Specifically,
finding 
\[
|M_\nu|_{ee}>m_{{\rm atm}}\qquad\hbox{and/or}\qquad \sum_\nu m_\nu >2 m_{%
{\rm atm}} 
\]
would prove the threefold degenerate spectrum. Furthermore, finding 
\[
0.01\,{\rm eV}<|M_\nu|_{ee}<m_{{\rm atm}} 
\]
and/or 
\[
\sum_\nu m_\nu>3|M_\nu|_{ee} 
\]
would exclude the SAMSW solution of the solar neutrino problem.
Finally, the only existence of a bound on $|M_\nu|_{ee}$, $|M_\nu|_{ee}<m_{%
{\rm atm}}$, together with SAMSW for solar neutrinos would prove the
`hierarchical' spectrum, since, for `degenerate' or `pseudo-Dirac'
neutrinos, the smallness of $\theta_{12}$ implies 
\[
|M_\nu|_{ee}\approx \max m_\nu \ge m_{{\rm atm}}. 
\]
In summary, disappointing as it may be, the experimental distinction
between the different neutrino spectra may be hard to achieve in absence of
a $0\nu,2\beta$ or a cosmological signal and with the solution of the solar
neutrino problem proven to be either LAMSW or VO. However, the VO solution of the
solar neutrino problem will be incompatible with a degenerate spectrum 
if supersymmetry with $\tan\beta\,\raise.3ex\hbox{$>$\kern-.75em\lower1ex%
\hbox{$\sim$}}\,10$ will be discovered. Of course it is also possible that
solar oscillations are not due to one of the three standard solutions and/or
that the LSND anomaly~\cite{LSND} will be confirmed.



\frenchspacing


\frenchspacing
\small\footnotesize
\begin{thebibliography}{nn}
\bibitem{sunexp}
\art{B.T. Cleveland}{\NP~(Proc. Suppl.)}{B38}{47}{1995};
\art[hep-ph/9812011]{SuperKamiokande collaboration}{Phys.Rev.Lett.}{82}{2430}{1999};
\art{K. Lande et al.}{Nucl.Phys.Proc.Suppl.}{77}{13}{1999};
\art{SAGE Collaboration}{Nucl.Phys.Proc.Suppl.}{77}{20}{1999};
\art{{\sc Gallex} collaboration}{Nucl.Phys.Proc.Suppl.}{77}{26}{1999}.


\bibitem{atmexp} 
\art[hep-ex/9805006]{Super-Kamiokande collaboration}{\PL}{B436}{33}{1998};
M. Messier, talk presented at the 1999 DPF meeting, Jan 1999,
available at the www address {\tt hep.bu.edu/\~{}messier/dpf/index.html}.
D. Casper, talk presented at the XXXIV Recontres de Moriond on Electroweak
interactions and unified theories, Les Arcs, 13--20/3/1999,
available at the www address 
{\tt moriond.in2p3.fr/EW/transparencies.}

\bibitem{sunfit}
For recent analyses of solar data see
\art[hep-ph/9807216]{J.N. Bahcall, P.I. Krastev and A. Yu Smirnov}{\PL}{B436}{243}{1998};
\hepart[hep-ph/9903262]{V. Barger and K. Whisnant};
\hepart[hep-ex/9903034]{M.B. Smy};
\hepart[hep-ph/9905458]{P.I. Krastev}. See also~\cite{lungo} for a non-standard analysis.

\bibitem{EL}
\hepart[hep-ph/9904279]{J. Ellis and S. Lola}.

\bibitem{madrid}
\hepart[hep-ph/9904395]{J.A. Casas, J.R. Espinosa, A. Ibarra and I. Navarro} and
{\em hep-ph/9905381}.


\bibitem{lungo}
\art[hep-ph/9807235]{R. Barbieri et al.}{\Jhep}{12}{017}{1998};

\bibitem{PD}
\art[hep-ph/9808333]{R. Barbieri, L. Hall and A. Strumia}{\PL}{B445}{407}{1999}.

\bibitem{deg}
\hepart[hep-ph/9901228]{R. Barbieri et al.}.

\bibitem{bimax}\hepart[hep-ph/9708483]{F. Vissani};
\hepart[hep-ph/9808293]{H. Georgi and S. Glashow}.

\bibitem{CEIN}
\hepart[hep-ph/9906281]{J.A. Casas, J.R. Espinosa, A. Ibarra and I. Navarro}.

\bibitem{Wu} \hepart[hep-ph/9905222]{Y-L. Wu}.


\bibitem{oka}
\hepart[hep-ph/9810471]{N. Haba, N. Okamura, M. Sugiura}.

\bibitem{CKP}
\hepart[hep-ph/9910231]{P. H. Chankowski, W. Kr\'olikowski and S. Pokorski}.


\bibitem{BW} For related considerations see
\hepart[hep-ph/9904281]{V. Barger and K. Whisnant}.

\bibitem{Lipari}
\hepart[hep-ph/9211320]{E. Akhedov, P. Lipari and M. Lusignoli}.

\bibitem{betabetaNow}
\hepart[hep-ex/9902014]{L. Baudis et al.}.

\bibitem{betabeta}
\art{H. Klapdor-Kleingrothaus, J. Hellmig and M. Hirsch}{J. Phys.}{G24}{483}{1998}.

\bibitem{cosm}
\art{W. Hu, D. Eisenstein and M. Regmark}{\PRL}{80}{5255}{1998}.

\bibitem{LSND}
\art{LSND collaboration}{\PR}{C54}{2685}{1996}
and
{\em Phys. Rev.} C58 (1998) 2489.

\end{thebibliography}
\end{document}